\documentclass[
reprint,
superscriptaddress,
frontmatterverbose, 
 amsmath,amssymb,
 aps,
prb,
]{revtex4-2}

\usepackage{graphicx}
\usepackage{dcolumn}
\usepackage{bm}
\usepackage{amsmath}
\usepackage{amssymb}
\usepackage{CJK}
\usepackage{keyval}
\usepackage{color}
\usepackage{txfonts}
\usepackage{verbatim}

\usepackage{lineno} 

\makeatletter

\newcommand{\Rmnum}[1]{\expandafter\@slowromancap\romannumeral #1@}
\makeatother

\begin{document}

\preprint{APS/123-QED}

\title{Observation of Electride-like $s$ States Coexisting with Correlated $d$ Electrons in NdNiO$_2$}

\author{Chihao Li}
\affiliation{Laboratory of Advanced Materials, State Key Laboratory of Surface Physics, \\and Department of Physics, Fudan University, Shanghai 200438, China}

\author{Yutong Chen}
\affiliation{Laboratory of Advanced Materials, State Key Laboratory of Surface Physics, \\and Department of Physics, Fudan University, Shanghai 200438, China}

\author{Xiang Ding}
\affiliation{Laboratory of Advanced Materials, State Key Laboratory of Surface Physics, \\and Department of Physics, Fudan University, Shanghai 200438, China}

\author{Yezhao Zhuang}
\affiliation{Shanghai Frontiers Science Research Base of Intelligent Optoelectronic and Perception, Institute of Optoelectronic and Department of Material Science, Fudan University, Shanghai 200433, China}

\author{Nan Guo}
\affiliation{Laboratory of Advanced Materials, State Key Laboratory of Surface Physics, \\and Department of Physics, Fudan University, Shanghai 200438, China}

\author{Zhihui Chen}
\affiliation{Laboratory of Advanced Materials, State Key Laboratory of Surface Physics, \\and Department of Physics, Fudan University, Shanghai 200438, China}

\author{Yu Fan}
\affiliation{Laboratory of Advanced Materials, State Key Laboratory of Surface Physics, \\and Department of Physics, Fudan University, Shanghai 200438, China}

\author{Jiahao Ye}
\affiliation{Laboratory of Advanced Materials, State Key Laboratory of Surface Physics, \\and Department of Physics, Fudan University, Shanghai 200438, China}

\author{Zhitong An}
\affiliation{Laboratory of Advanced Materials, State Key Laboratory of Surface Physics, \\and Department of Physics, Fudan University, Shanghai 200438, China}

\author{Suppanut Sangphet}
\affiliation{Laboratory of Advanced Materials, State Key Laboratory of Surface Physics, \\and Department of Physics, Fudan University, Shanghai 200438, China}

\author{Shenglin Tang}
\affiliation{Laboratory of Advanced Materials, State Key Laboratory of Surface Physics, \\and Department of Physics, Fudan University, Shanghai 200438, China}

\author{Xiaoxiao Wang}
\affiliation{Laboratory of Advanced Materials, State Key Laboratory of Surface Physics, \\and Department of Physics, Fudan University, Shanghai 200438, China}

\author{Hai Huang}
\affiliation{Shanghai Frontiers Science Research Base of Intelligent Optoelectronic and Perception, Institute of Optoelectronic and Department of Material Science, Fudan University, Shanghai 200433, China}

\author{Haichao Xu}
\email{xuhaichao@fudan.edu.cn}
\affiliation{Laboratory of Advanced Materials, State Key Laboratory of Surface Physics, \\and Department of Physics, Fudan University, Shanghai 200438, China}
\affiliation{Shanghai Research Center for Quantum Sciences, Shanghai 201315, China}

\author{Donglai Feng}
\email{dlfeng@hfnl.cn}
\affiliation{New Cornerstone Science Laboratory, Hefei National Laboratory, Hefei, 230026, China}

\author{Rui Peng}
\email{pengrui@fudan.edu.cn}
\affiliation{Laboratory of Advanced Materials, State Key Laboratory of Surface Physics, \\and Department of Physics, Fudan University, Shanghai 200438, China}
\affiliation{Shanghai Research Center for Quantum Sciences, Shanghai 201315, China}

\date{\today}

\begin{abstract}
Despite exhibiting a similar $d_{x^2-y^2}$ band character to cuprates, infinite-layer nickelates host additional electron pockets that distinguish them from single-band cuprates. The elusive orbital origin of these electron pockets has led to competing theoretical scenarios.
Here, using polarization-dependent and resonant angle-resolved photoemission spectroscopy (ARPES), we determine the orbital character of the Fermi surfaces in NdNiO$_2$.
Our data reveal that the electron-like pocket arises predominantly from interstitial $s$ states, with negligible contributions from rare-earth 5$d$ and 4$f$ orbitals near the Fermi level. The observation of well-defined quantum well states indicates a uniform distribution of these interstitial electrons throughout the film thickness. 
By comparing with electronic structure of LaNiO$_2$, we find that the rare-earth element modulates the Ni-derived bands and hopping integrals through a chemical pressure effect. These findings clarify the role of rare-earth elements in shaping the low-energy electronic structure
and establish the presence of electride-like interstitial $s$ states in a correlated oxide system, where electrons occupy lattice voids rather than atomic orbitals.
The electride-like character offer new insight into the self-doping and superconductivity in infinite-layer nickelates.

\end{abstract}

\maketitle

$Introduction-$Infinite-layer nickelates have been proposed as analogs to cuprate superconductors due to their square-planar NiO$_2$ motif and nominal Ni$^{1+}$ 3$d^9$ valence configuration \cite{anisimov1999electronic}. 
Recent angle-resolved photoemission spectroscopy (ARPES) studies have revealed a hole-like Fermi pocket with Ni 3$d_{x^2-y^2}$ character in both LaNiO$_2$\cite{ding2024cuprate} and its doped counterparts \cite{ding2024cuprate,sun2024electronic}, closely resembling cuprates. However, additional electron pockets near the A point have been observed \cite{ding2024cuprate,sun2024electronic}, revealing multi-band Fermi surfaces distinct form the single-band cuprates.
The orbital origin of these electron pockets remains under active debate, giving rise to competing theoretical scenarios\cite{si2024closing,labollita2023conductivity,lechermann2020multiorbital,karp2020comparative,zhang2020self,lee2004infinite,foyevtsova2023distinct,gu2020substantial,nomura2019formation,wu2020robust,adhikary2020orbital,worm2024spin,li2022two,zhang2021magnetic,zhang2021magnetic,jiang2019electronic,choi2020role,bandyopadhyay2020superconductivity}. 
Furthermore, some theoretical studies propose that hybridization between rare-earth ($RE$) 4$f$ electrons and Ni 3$d$ or $RE$ 5$d$ orbitals reshapes the Fermi surfaces, introducing flat 4$f$ bands near the Fermi level \cite{zhang2021magnetic,jiang2019electronic,choi2020role,bandyopadhyay2020superconductivity}. 
This possible involvement of $RE$ 5$d$ and 4$f$ orbitals contrasts with the situation in cuprates and iron-based superconductors, where the Fermi surfaces are derived purely from transition-metal 3$d$ orbitals hybridized with ligand $p$ states \cite{kamihara2008iron,stewart2011superconductivity}.
Despite extensive theoretical efforts \cite{si2024closing,labollita2023conductivity,lechermann2020multiorbital,karp2020comparative,zhang2020self,lee2004infinite,foyevtsova2023distinct,gu2020substantial,nomura2019formation,wu2020robust,adhikary2020orbital,worm2024spin,li2022two,zhang2021magnetic,zhang2021magnetic,jiang2019electronic,choi2020role,bandyopadhyay2020superconductivity}, the orbital composition at the Fermi energy remains  experimentally unresolved, hindering a consistent understanding of superconductivity in nickelates.  

In prevailing theories, $RE$ 5$d$ orbitals are predicted to  contribute directly at the  Fermi level \cite{si2024closing,labollita2023conductivity,lechermann2020multiorbital,karp2020comparative,zhang2020self,lee2004infinite}, potentially hybridizing with Ni 3$d$ electrons to form Kondo singlets \cite{zhang2020self,labollita2023conductivity,wu2020robust,si2024closing,adhikary2020orbital}, and even hosting enhanced electron-phonon coupling (EPC) that may dominate superconductivity \cite{li2022two}. 
An alternative scenario suggests contribution from interstitial $s$-like electrons residing in lattice voids created by the absence of apical oxygen \cite{foyevtsova2023distinct,gu2020substantial,si2024closing,nomura2019formation}, forming delocalized states not centered on any atomic site, reminiscent of electride materials. While such states are established in weakly correlated systems such as [Ca$_{24}$Al$_{28}$O$_{64}$]$^{4+}$(4e$^-$), {[Ca$_2$N]$^{+}$(e$^-$)} and {[Y$_2$C]$^{2+}$(2e$^-$)} \cite{matsuishi2003high,lee2013dicalcium,hirayama2018electrides}, their existence in correlated transition-metal oxides remains purely theoretical. Since the interstitial electrons may hybridize with Ni 3$d$ states, experimental verification of such electride-like behavior may offer new perspectives for tuning superconductivity. 
Moreover, recent experiments show that the optimal superconducting transition temperature ($T_c$) increases systematically with rare-earth substitution toward heavier $RE$ elements \cite{osada2021nickelate,zeng2022superconductivity,osada2020phase,li2019superconductivity,lee2023linear,wei2023superconducting,lee2024millimeter,chow2025bulk}, reaching nearly 40~K in latest reports \cite{chow2025bulk,yang2025enhanced}. This highlights the critical role of rare-earth elements in both electronic structure and superconductivity. 
To clarify the pairing mechanism of infinite-layer nicklates, it is therefore crucial to resolve the orbital composition of Fermi surfaces and understand how rare-earth substitution modifies the electronic structure. Here we perform polarization dependent and resonant ARPES studies on NdNiO$_2$, and compare its electronic structure with that of LaNiO$_2$.

\begin{figure*}[t]
\includegraphics[width=170mm]{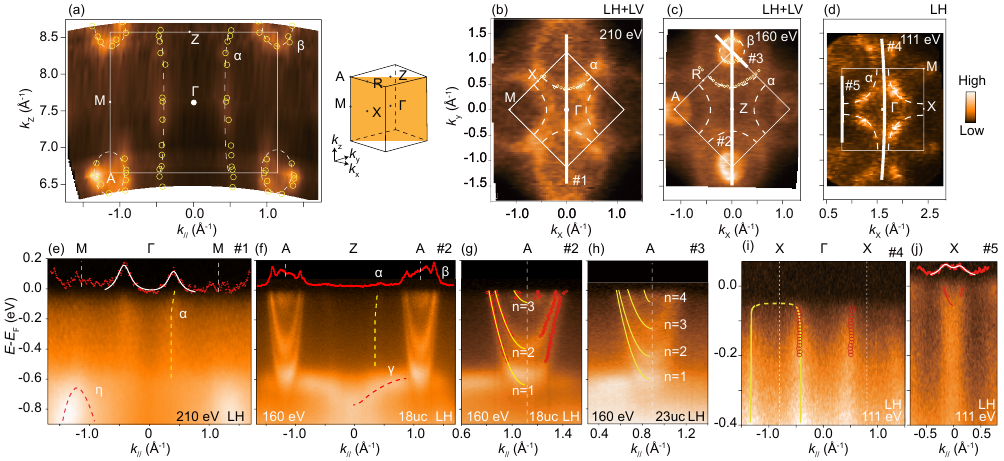}
\caption{\label{Figure2} 
(a) Photoemission intensity map in the $\Gamma$-M-A-Z plane 
integrated over an energy window of [$E_{\rm{F}}$-50~meV,$E_{\rm{F}}$+50~meV]. The inner potential is estimated to be 17~eV. 
(b) Photoemission intensity map at the $\Gamma$ plane over an energy window of [$E_{\rm{F}}$-50~meV,$E_{\rm{F}}$+50~meV] combining both LH and LV polarizations. 
(c) Same as panel~b but at the Z plane. 
(d) Same as panel~b but measured on the sample rotated 45$^\circ$ in plane. 
(e-f) Photoemission intensity  along M-$\Gamma$-M (e) and  along A-Z-A (f).
The momentum distribution curves (MDCs) at $E_{\rm{F}}$ are overlaid in panels e and f to show the Fermi crossings.
(g-h) Photoemission spectrum around the A point of the 18~uc NdNiO$_2$ films (g) and  on a 23~uc NdNiO$_2$ films (h), showing three and four subbands, respectively.
(i) Photoemission intensity along $\Gamma$~-~X (i) and  along $\Gamma$~-~M (j).  The MDC at $E_{\rm{F}}$ is overlaid to show the Fermi crossings.}
\end{figure*}

Performing ARPES measurements on NdNiO$_2$ presents substantial experimental challenges. A primary difficulty lies in synthesizing homogeneous, single-crystalline infinite-layer films with phase purity maintained both at the surface and in the bulk \cite{parzyck2024absence, parzyck2024synthesis}.
Incomplete removal of apical oxygen during the reduction process often leads to residual oxygen incorporation  \cite{parzyck2024superconductivity, parzyck2024absence, WOS:001338349500001, raji2023charge}, which can induce a 3$a_0$ superlattice modulation observable in resonant elastic X-ray scattering (REXS) measurements \cite{tam2022charge, krieger2022charge,parzyck2024superconductivity, parzyck2024absence, WOS:001338349500001, raji2023charge}.
The absence of such superlattice signals has been demonstrated only in optimized SrTiO$_3$-capped NdNiO$_2$/SrTiO$_3$ films \cite{parzyck2024absence,raji2023charge,rossi2024universal}. However, these capped films are unsuitable for vacuum ultraviolet (VUV) ARPES due to the limited probing depth of $\sim$1-2 unit cells, which hinders access to the underlying electronic structure.
These challenges underscore the necessity of preparing uncapped NdNiO$_2$ that are fully-reduced  and without 3$a_0$ reconstruction in order to access the intrinsic electronic structure.
A further difficulty arises from maintaining atomically flat, single-crystalline surfaces after the topotactic reduction process. 
We overcome these challenges by synthesizing high quality, uncapped NdNiO$_2$/SrTiO$_3$ films with well-ordered surface under $in$-$situ$conditions \cite{ding2024cuprate} (see Supplementary Fig.~S1 for detailed characterizations). REXS measurements confirm the absence of residual-oxygen-induced superlattice modulations, indicating complete reduction of the films [see Supplementary Section \Rmnum{1} for details].

\begin{figure*}[htbp]
\includegraphics[width=170mm]{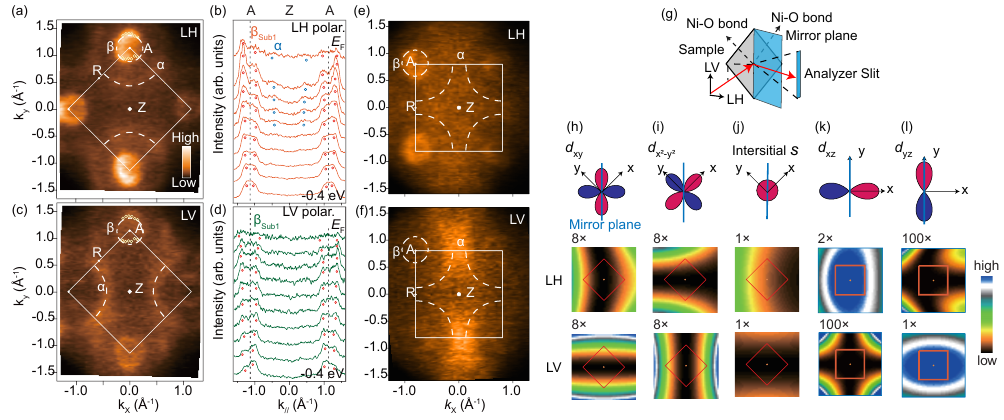}
\caption{\label{Figure3} 
(a) Photoemission intensity map of NdNiO$_2$ at the Z-R-A plane, integrated over an energy window of [$E_{\rm{F}}$-50~meV,~$E_{\rm{F}}$+50~meV] measured in LH polarization. 
(b) MDCs along A-Z-A direction measured in LH polarization.
(c) Same as panel~a but in LV polarization.
(d) Same as panel~b but in LV polarization. 
(e) Same as panel~a but measured on the sample rotated 45$^\circ$ in plane. 
(f) Same as panel~e but in LV polarization.
(g) Experimental geometry of the polarization-dependent ARPES measurements. 
(h-l) Illustration of the spatial symmetry of $d_{\rm{xy}}$, $d_{\rm{x^2-y^2}}$, the interstitial $s$, $d_{\rm{xz}}$ and $d_{\rm{yz}}$ orbitals with respect to the mirror plane determined by the analyzer slit (along A-Z-A).  Simulated photoemission intensity of atomic orbitals well captures the symmetry analysis after considering the out-of-plane components (adapted from ref.~\onlinecite{ye2013angle}). 
}
\end{figure*}

$Results-$ The Fermi surfaces of NdNiO$_2$ in the out-of-plane $\Gamma$-M-A-Z plane consists of a small  $\beta$ pocket and a large $\alpha$ pocket around the Brillouin zone corner [Fig.~\ref{Figure2}(a)]. 
The $\beta$ band shows electron-like dispersion, and locate only around the A point in the Z-R-A plane [Figs.~\ref{Figure2}(c) and \ref{Figure2}(f)], but is absent at the $\Gamma$-X-M plane [Figs.~\ref{Figure2}(b) - \ref{Figure2}(e)], demonstrating its three-dimensional character.
The $\alpha$ band show quasi-two-dimensional character, whose dispersion forms a large hole-like pocket centered at zone corner in both $\Gamma$-X-M plane [Figs.~\ref{Figure2}(b) and \ref{Figure2}(e)] and Z-R-A plane [Figs.~\ref{Figure2}(c) and \ref{Figure2}(f)]. 
The electronic structure of NdNiO$_2$ closely resembles that of LaNiO$_2$ \cite{ding2024cuprate}, with a similarly sized $\alpha$ pocket that appears more rounded in NdNiO$_2$ and more square-like in LaNiO$_2$ (see Supplementary Fig.~S3).
Besides, the identical size of $\beta$ pocket indicates a similar self-doping level (see Supplementary Fig.~S4).
The observed three-dimensional Fermi surface topology exhibits good agreement with the bulk Brillouin zone of the NdNiO$_2$ lattice [Fig.~\ref{Figure2}(a)], reinforcing that the ARPES spectra capture the intrinsic bulk electronic structure of NdNiO$_2$.

In the 18-unit-cell (uc) NdNiO$_2$ films, two additional sub-bands appear near the A point with lower binding energies than the $\beta$ band, crossing the Fermi level and contributing to the spectral weight inside the $\beta$ pocket [Figs.\ref{Figure2}(a), \ref{Figure2}(c), and \ref{Figure2}(f)–\ref{Figure2}(g)]. In thicker 23uc films, a third sub-band emerges, resulting in a total of four sub-bands including the outermost $\beta$ band [Fig.~\ref{Figure2}(h)].
The Fermi crossing size and the pocket shape of the outermost $\beta$ band remain intact (Supplementary Fig.~S5). 
The thickness dependence of the sub-band number fits with the characteristics of quantum well states \cite{gauthier2020quantum,kobayashi2015origin,kawasaki2018engineering,kawakami1999quantum}, which arises from the confinement of the electronic wave functions within the thickness of the film (see Supplementary Section \Rmnum{7} for detailed discussion). 
Our observation of exceptionally well-resolved quantum-well states demonstrates both the high quality of the NdNiO$_2$ films and homogeneous electronic states throughout the film thickness.

No electron pocket is observed at the $\Gamma$ point in our measurements [Fig.~\ref{Figure2}(b)], disfavoring theoretical predictions of a pocket composed of hybridized rare-earth 5$d_{3z^2-r^2}$ and Ni 3$d_{3z^2-r^2}$ orbitals \cite{labollita2023conductivity, wu2020robust, li2022two, gu2020substantial, zhang2021magnetic}. A similar absence of a $\Gamma$-point electron pocket has also been reported in LaNiO$_2$ \cite{ding2024cuprate}, suggesting a shared characteristic across different rare-earth variants.
The absence of the pocket persists across all measurement geometries [including 45$^\circ$ in plane rotation as shown in Fig.~\ref{Figure2}(d)] and temperature ranges spanning the resistivity upturn (Supplementary Section \Rmnum{8}), ruling out the suppression by matrix element effects or magnetic transition scenarios \cite{jiang2023variation}. 
All bands are found to lie below the Fermi level along cut \#4 [Fig.~\ref{Figure2}(d)].
The hole-like band top observed along cut $\#$4 [Fig.~\ref{Figure2}(i)] matches the electron-like band bottom along cut $\#$5 [Fig.~\ref{Figure2}(j)], consistently indicating that the only observed band is the $\alpha$ band, which forms a saddle point near the X point.
Such saddle point behavior is similar to the the $d_{{x^2-y^2}}$ band observed in cuprates and LaNiO$_2$ family \cite{damascelli2003angle,ding2024cuprate}.

\begin{figure*}[htbp]
\includegraphics[width=170mm]{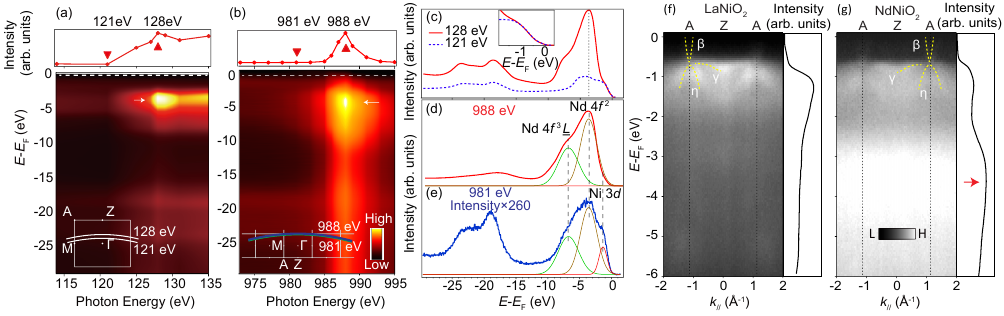}
\caption{\label{Figure4}
(a) Momentum-integrated photoemission intensity of NdNiO$_2$ as a function of photon energy across the Nd $N_4$ edge ($4d\rightarrow4f$). The upper inset displays the photon-energy dependence of the total intensity integrated over a 30 eV energy window. The lower inset shows the corresponding $k_z$ locations with photon energy of 121~eV and 128~eV near the M-$\Gamma$-M direction. 
(b) Same as panel~a, but measured across the Nd $M_5$ edge ($3d_{5/2}\rightarrow4f$). 
(c) Momentum-integrated photoemission intensity of NdNiO$_2$ taken by photon energies of 128 eV and 121 eV, respectively. Inset: enlarged view near $E_{\rm{F}}$.
(d-e) Momentum-integrated photoemission intensity of NdNiO$_2$ taken with 988~eV (on-resonance) and 981~eV (on-resonance) photons, respectively. The peak positions were obtained by Gaussian fittings.
(f-g) Photoemission intensity of LaNiO$_2$/SrTiO$_3$ and NdNiO$_2$/SrTiO$_3$ along A-Z-A direction, respectively. 
}
\end{figure*}

The orbital components of Fermi surfaces are studied and analyzed using polarization-dependent ARPES measurements (Fig.~\ref{Figure3}). 
The $d_{x^2-y^2}$ orbital, being odd with respect to the mirror plane in our experimental geometry [Fig.~\ref{Figure3}(g)], is expected to be visible under linearly horizontal (LH) polarization and suppressed under linearly vertical (LV) polarization along the vertical A–Z–A cut [Fig.~\ref{Figure3}(i)]. This polarization-dependent behavior is clearly observed for the $\alpha$ band [Figs.~\ref{Figure3}(a)–\ref{Figure3}(d)], confirming its $d_{x^2-y^2}$ orbital character, which is consistent with previously reported in LaNiO$_2$ \cite{ding2024cuprate}.
As for the $\beta$ band, 
theoretical calculations have proposed possible contributions from Nd 5$d_{xy}$ orbitals \cite{wu2020robust,adhikary2020orbital,nomura2019formation,worm2024spin}, interstitial $s$ states (also named zeronium) \cite{si2024closing,foyevtsova2023distinct,gu2020substantial} and Ni $d_{xz}$/$d_{yz}$ orbitals \cite{foyevtsova2023distinct,si2024closing}. 
Note that $d_{xy}$ orbital should be suppressed along the A–Z–A direction under LH polarization according to the photoemission selection rules [Fig.~\ref{Figure3}(h)]. However, the $\beta$ band exhibits strong intensity under LH polarization along this direction [Figs.~\ref{Figure3}(a)–\ref{Figure3}(b)], disfavoring a dominant contribution from $d_{xy}$ orbitals.
As shown by ARPES measurements on the NdNiO$_2$ sample with a 45-degree in-plane rotation [Figs.~\ref{Figure3}(e)-\ref{Figure3}(f)], $\beta$ pocket is significantly suppressed under LV polarization, disfavoring the $d_{yz}$ orbital which would be significantly enhance by LV polarization [Fig.~\ref{Figure3}(l)]. As the in-plane structure of NdNiO$_2$ is four-fold symmetric, if $d_{yz}$ is absent, $d_{xz}$ should be absent by symmetry.
These results exclude the $d_{xz}$/$d_{yz}$ component for the $\beta$ band.
Instead, the consistently strong intensity of the $\beta$ pocket under LH polarization regardless of in-plane rotation [Figs.~\ref{Figure3}(a) and \ref{Figure3}(e)] agrees with the dominant contribution from interstitial $s$ states, which are isotropic and have out-of-plane character that enhances the overall intensity under LH polarization [Fig.~\ref{Figure3}(h)]. 
These results establish that the $\alpha$ band is primarily composed of Ni 3$d_{x^2-y^2}$ orbitals and, more significantly, identify the $\beta$ band as originating from interstitial $s$ states, after systematically excluding alternative orbital assignments.

To trace the contribution from Nd, we further perform resonant photoemission measurements across the Nd absorption edge.
A resonance enhancement of integrated photoemission intensity emerges above a photon energy of 121~eV and reaches a maximum at 128~eV [Fig.~\ref{Figure4}(a)], and above a photon energy of 985~eV and reaches a maximum at 988~eV [Fig.~\ref{Figure4}(b)], corresponding to Nd $N_4$-edge ($4d \rightarrow 4f$) and Nd $M_5$-edge ($3d_{5/2} \rightarrow 4f$), respectively.
Notably, the measurements at 128~eV and 988~eV locate the $k_z$ near the $\Gamma$-X-M plane [inset of Fig.~\ref{Figure4}(a)] the Z–A–R plane in $k_z$ [inset of Fig.~\ref{Figure4}(b)], respectively, and thus can determine the Nd contribution at all the bands near $E_F$.
The most prominent enhancement is observed in valence states at $E_{\rm{F}}$-8 eV and $E_{\rm{F}}$-3.4 eV  at both resonance energies [Figs.~\ref{Figure4}(c)-\ref{Figure4}(d)]. 
In contrast, no resonant enhancement is observed within -1~eV to $E_{\rm{F}}$ [inset of Fig.~\ref{Figure4}(c)], indicating that Nd states, especially the Nd 4$f$ states, are undetectably weak in this energy range at both the $\Gamma$-X-M and Z-R-A plane. 
The energy position and resonance behavior of the enhanced states at  $E_{\rm{F}}$-3.4eV and $E_{\rm{F}}$-8eV resemble those seen in Nd$_{2-x}$Ce${_x}$CuO${_4}$, where they have been attributed to the Nd 4$f^{2}$ final state and the 4$f^3\underline{L}$ final state, respectively \cite{namatame1990resonant,fujimori19884f}. 
Therefore, Nd 4$f$ states are localized and make negligible contributions to both the Fermi surfaces and the formation of the low-lying $\alpha$ and $\beta$ bands.

\begin{figure}[tb]
\includegraphics[width=86mm]{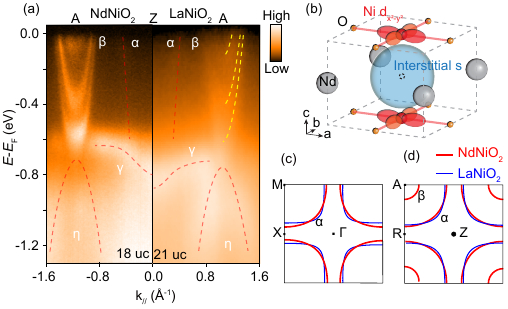}
\caption{ \label{Figure5}
(a) Comparison of photoemission intensity between NdNiO$_2$/SrTiO$_3$ and LaNiO$_2$/SrTiO$_3$ near $E_{\rm{F}}$ along A-Z-A direction. 
(b) Illustration of the atomic unit cell of NdNiO$_2$, together with the Ni 3$d_{x^2-y^2}$ and interstitial $s$ orbitals. 
(c-d) Comparison of the measured Fermi surfaces of NdNiO$_2$ and LaNiO$_2$ in the $\Gamma$-X-M plane and Z-R-A plane, respectively.
}
\end{figure}

By comparing the electronic structures of NdNiO$_2$ and LaNiO$_2$, we find that the valence bands are largely similar [Figs.~\ref{Figure4}(f)–\ref{Figure4}(g)], except for an additional flat band at approximately $E_{\rm{F}}$–3.4 eV in NdNiO$_2$, which corresponds to the resonantly enhanced spectral weight at the Nd $N$-edge. Since La lacks 4$f$ electrons, this feature confirms its origin from localized Nd 4$f$ orbitals. These 4$f$ states do not contribute to the Fermi surface and are unlikely to participate in superconducting pairing, but may give rise to localized magnetic moments, possibly accounting for the distinct magnetoresistance symmetry behaviors observed between NdNiO$_2$ and LaNiO$_2$ \cite{wang2023effects}.
Near $E_{\rm{F}}$, both NdNiO$_2$ and LaNiO$_2$ exhibit similar band dispersions and Fermi crossings [Fig.~4(a) and Fig.~S4(e)], 
highlighting a common low-energy electronic structure across the infinite-layer nickelates, where the Fermi surface is primarily governed by correlated Ni 3$d_{x^2-y^2}$ states and itinerate interstitial $s$ orbitals [Fig.~4(b)].
Nevertheless, certain differences can be identified between NdNiO$_2$ and LaNiO$_2$. 
The flat $\gamma$ band below $E_{\rm{F}}$, attributed to Ni 3$d_{3z^2-r^2}$ character based on its strong LH polarization response (Supplementary Fig.~S8), shifts to lower binding energy in NdNiO$_2$ [Fig.~\ref{Figure5}(a)].
This shift likely reflects a chemical pressure effect induced by the smaller lattice constant of NdNiO$_2$ (3.285~\AA) relative to LaNiO$_2$ (3.41~\AA) \cite{ding2024cuprate}.
The Ni $d_{x^2-y^2}$-derived $\alpha$ pocket also shows a more rounded Fermi surface in NdNiO$_2$, as reflected in the Fermi crossing positions [Figs.~\ref{Figure4}(c)-(d) and  Fig.~S4]. These differences suggest that rare-earth substitution tunes the Ni orbital energies and hopping parameters via chemical pressure effects.

$Discussion-$The absence of Nd orbital contributions at $E_{\rm{F}}$ strongly disfavors scenarios where rare-earth 5$d$ or 4$f$ states directly participate in superconductivity. Some low-lying Nd-derived states may be pushed above $E_{\rm{F}}$ due to crystal field splitting \cite{kugler2024low}, where they can still contribute to the 0.6–0.7~eV excitation peak observed in RIXS studies via a charge-transfer process \cite{hepting2020electronic,lu2021magnetic}. 
Furthermore, the $\beta$ pocket is relatively small, and the corresponding low density of states appears insufficient to support a dominant BCS-type superconducting mechanism proposed by GW calculations \cite{li2022two}.

The identification of interstitial $s$ electrons in NdNiO$_2$ marks a significant extension of the electride concept into strongly correlated oxides. These delocalized carriers, occupying lattice voids rather than atomic orbitals, coexist with the correlated Ni 3$d_{x^2-y^2}$ states and jointly shape the Fermi surfaces. 
Located between adjacent Ni atoms along $c$ [Fig.~\ref{Figure5}(b)], the interstitial $s$ states can naturally hybridize with both Ni 3$d_{3z^2-r^2}$ \cite{foyevtsova2023distinct} and, via oxygen-mediated pathways, Ni 3$d_{x^2-y^2}$ orbitals \cite{gu2020substantial}.
The presence of these electride-like interstitial electrons may further endow nickelates with exotic characters such as low work function, high carrier mobilities, and free-electron-like behavior \cite{hirayama2018electrides}.
These findings call for theoretical frameworks that explicitly incorporate interstitial degrees of freedom in modeling the electronic structure and pairing interactions in infinite-layer nickelates.

Although rare-earth atomic orbitals do not directly participate at the Fermi surfaces, the rare-earth element likely serve as chemical pressure tuning knob \cite{sharma2024pressure,bernardini2022geometric}.
The contraction of $c$ lattice in NdNiO$_2$ shifts the Ni 3$d_{3z^2-r^2}$ band to lower binding energy.
The Ni–Ni hopping parameters are also modified by Nd substitution, leading to the change of Fermi surface shape following the tight-binding models of cuprate-like systems \cite{maier2020overdoped,wu2018pseudogap,honerkamp2001magnetic}.
The structural and electronic modifications may be key to the enhanced $T_{\rm{c}}$  under physical pressure \cite{wang2022pressure,di2024unconventional} and  rare-earth substitution, such as with Sm \cite{chow2025bulk}.

$Conclusion-$Using optimized thin-film synthesis, resonant photoemission, and polarization-dependent ARPES, we have identified Ni 3$d_{x^2-y^2}$ and interstitial $s$ states as the dominant contributors to the Fermi surface in infinite-layer nickelates. These results resolve the orbital composition of the low-energy electronic structure and provide direct evidence for electride-like states in a correlated oxide system. The interplay between spatially delocalized interstitial carriers with correlated Ni 3$d$ electrons may play a critical role in the superconductivity of infinite-layer nickleates.
Rare-earth elements contribute minimally at the Fermi surface but modulate the electronic structure via chemical pressure effects, tuning the  bandwidth and hopping amplitudes. 
Our results impose important experimental constraints for theoretical models and offer new insight into the minimal ingredients required for superconductivity in infinite-layer nickelates.

\begin{acknowledgments}

We thank G. M. Zhang, D. F. Li, X. Zhang and Q. Wu for helpful discussions, and Z. T. Liu, J. S. Liu, Y. Mao, J. Y. Ding, J. Y. Liu, Y. W. Cheng, Y. B. Huang, Z. H. Chen for experimental support. This work was supported by the National Natural Science Foundation of China (Grants Nos. 92477206, 12422404, 12274085, 12204109), the National Key R\&D Program of China (2023YFA1406300), the New Cornerstone Science Foundation, the Innovation Program for Quantum Science and Technology (2021ZD0302803), the Shanghai Municipal Science and Technology Major Project (2019SHZDZX01), the Shanghai Science and Technology Commission (21JC1400200), and the China National Postdoctoral Program for Innovative Talents (BX20230078). ARPES measurements were performed at BL03U and BL09U of the Shanghai Synchrotron Radiation Facility.
REXS measurements were performed at SSRL beamline 13-3, SLAC National Accelerator Laboratory.
\end{acknowledgments}
\appendix
\bibliography{apssamp}
\end{document}